\documentclass[aps,prb,twocolumn,superscriptaddress]{revtex4-1}

\usepackage{amsmath,amsthm,amssymb}
\usepackage[dvipdfmx]{graphicx}
\usepackage{amsmath,bm}
\usepackage{color,ulem}
\newcommand{\1}{\mbox{1}\hspace{-0.25em}\mbox{l}}

\def\n{\nonumber \\ }

\let\olditemize\itemize
\renewcommand{\itemize}{
   \olditemize
   \setlength{\itemsep}{1pt}
   \setlength{\parskip}{0pt}
   \setlength{\parsep}{0pt}
}

\newlength{\figwidth}
\newlength{\figlarge}
\begin{document}
\title{Bosonic symmetry protected topological phases with reflection symmetry}

\author{Tsuneya Yoshida} 
\affiliation{Condensed Matter Theory Laboratory, RIKEN, Wako, Saitama, 351-0198, Japan}
\author{Takahiro Morimoto} 
\altaffiliation[Current address: ]{Department of Physics, University of California, Berkeley, CA 94720}
\affiliation{Condensed Matter Theory Laboratory, RIKEN, Wako, Saitama, 351-0198, Japan}
\affiliation{RIKEN Center for Emergent Matter Science (CEMS), Wako, Saitama, 351-0198, Japan}
\author{Akira Furusaki} 
\affiliation{Condensed Matter Theory Laboratory, RIKEN, Wako, Saitama, 351-0198, Japan}
\affiliation{RIKEN Center for Emergent Matter Science (CEMS), Wako, Saitama, 351-0198, Japan}
\date{\today}
\begin{abstract}
We study two-dimensional bosonic symmetry protected
topological (SPT) phases which are protected by reflection symmetry
and local symmetry
[$Z_N\rtimes R$, $Z_N\times R$, U(1)$\rtimes R$, or U(1)$\times R$],
in the search for two-dimensional bosonic analogs of topological crystalline
insulators in integer-$S$ spin systems with reflection and spin-rotation
symmetries.
To classify them, we employ a Chern-Simons approach and examine
the stability of edge states against perturbations that preserve
the assumed symmetries.
We find that SPT phases protected by
$Z_N\rtimes R$ symmetry are classified as
$\mathbb{Z}_2\times\mathbb{Z}_2$ for even $N$
and 0 (no SPT phase) for odd $N$
while those protected by U(1)$\rtimes R$ symmetry are $\mathbb{Z}_2$.
We point out that the two-dimensional Affleck-Kennedy-Lieb-Tasaki state
of $S=2$ spins on the square lattice is a $\mathbb{Z}_2$ SPT phase
protected by reflection and $\pi$-rotation symmetries.
\end{abstract}
\maketitle



\section{Introduction}
Topological insulators and superconductors are gapped phases of
noninteracting fermions in which the ground state wave functions have
topologically nontrivial structures in the presence of certain symmetry
constraints.
The topological structures in the bulk wave functions guarantee
the presence of gapless excitations at the boundary which are robust
against any perturbation respecting the symmetry constraints.
These gapless excitations lead to novel transport properties. 
For example, time-reversal-invariant topological insulators
in three dimensions are characterized
by $\mathbb{Z}_2$ topological invariants in the bulk.\cite{TI_review_Hasan10,TI_review_Qi10}
Correspondingly, stable gapless Dirac fermions emerge at the surface,
which lead to the topological magnetoelectric effect.\cite{Qi_3DTI2008,Essin_TME2009,Essin_TME2010}
Topological insulators and superconductors of noninteracting fermions
have been classified, in terms of the time-reversal, particle-hole,
and chiral symmetries, into the ``periodic table."\cite{Schnyder_classification_free_2008,Kitaev_classification_free_2009,Ryu_classification_free_2010} 

Interacting bosons can also host gapped phases with
gapless boundary excitations that are stable against any perturbation
as far as certain symmetries are preserved.
These phases are dubbed bosonic symmetry protected topological (SPT) phases.
One of representative examples in one dimension is the Haldane phase
in the spin-1 antiferromagnetic Heisenberg chain.\cite{Haldane_Haldanephase_1983,AKLT_model1987}
At each end of the Haldane spin chain emerge zero-energy spin-$\frac12$
degrees of freedom which are protected by
$\pi$-rotation about two orthogonal axes in the spin space
or by the time-reversal symmetry. \cite{entanglement_Pollmann10}
Bosonic SPT phases in two dimensions are also studied in two-component
bose gas with U(1) symmetry.\cite{senthil_IQHE_2013,Wu_IQHE_2013,Furukawa_IQHE_2013,Regnault_IQHE_2013}
The SPT phases with local symmetries have been classified with
various methods
including matrix product state representation,\cite{entanglement_Pollmann10,Turner11,Fidkowski_1Dclassificatin_11}
group cohomology,\cite{Chen_classification_1D_1,Chen_classification_1D_1,chen_cohomology_3D,gu_supercohomology}
Chern-Simons theory,\cite{Neupert_CS_2011,Lu_CS_2011,Levin_CS_2012,Hsieh_CS_CPT_2014}
nonlinear sigma models,\cite{Cenke2014}
and cobordisms.\cite{kapustin_bosonic_cobordisms2014_1,kapustin_bosonic_cobordisms2014_2,kapustin_fermionic_cobordisms2014}

Recently, the concept of topological phases of noninteracting fermions
has been further extended to topological crystalline insulators by including
spatial symmetries.\cite{Fu_TCI_2011}
Experimental realization of topological crystalline insulator with
reflection symmetry is achieved
in SnTe.\cite{Tanaka_TCI_SnTe2012,Hsieh_TCH_SnTe_2012} 
Surfaces of this material have an even number of Dirac cones,
and the strong $\mathbb{Z}_2$ index for topological insulators is trivial.
Thus, the time-reversal symmetry does not protect these Dirac cones.
Instead, reflection symmetry is responsible for the symmetry protection
of Dirac cones, allowing for a nontrivial mirror Chern number defined
on the reflection invariant plane in the three-dimensional Brillouin zone.
Now, a natural question we may ask is whether analogs of topological
crystalline insulators of fermions exist in bosonic SPT phases.
The bosonic SPT phases protected by spatial symmetries have not been
fully explored so far, while there are several recent attempts
in terms of their classification.\cite{Cenke2014,Hsieh_PRB14_01,Hsieh_CS_CPT_2014,Ware_TCI_chain_2015,Hermele_bosonicTCI_2015}
In particular, few realistic models are known for such two-dimensional (2D)
SPT phases with spatial symmetries.

Motivated by these, we explore topological crystalline insulators
in interacting bosonic systems.
Specifically, focusing on integer-spin systems, we study 2D
bosonic SPT phases protected by reflection symmetry as well as
spin-rotation symmetry. 
In order to classify SPT phases, we employ the Chern-Simons
approach\cite{Lu_CS_2011} which is well suited for classifying 2D
SPT phases with reflection symmetry.
In this approach, SPT phases are classified by analyzing the stability
of gapless edge modes against any perturbation allowed under given symmetry
constraints.
If all gapless modes are gapped out without symmetry breaking,
the system is in a trivial phase; otherwise it is an SPT phase. 
We apply this method to integer-spin systems protected by reflection
and discrete spin-rotation symmetry ($Z_N\rtimes R$). 
Our classification results show that SPT phases form $\mathbb{Z}_2\times\mathbb{Z}_2$ group
for even $N$, while no SPT phase is allowed for odd $N$.
Furthermore, the Affleck-Kennedy-Lieb-Tasaki (AKLT) state of $S=2$ spins
on the square lattice is shown to be a bosonic SPT phase characterized by
a $\mathbb{Z}_2$ invariant.
This can be regarded as a topological crystalline insulator in spin systems.

The rest of this paper is organized as follows. 
Section~\ref{sec: framework} summarizes the classification scheme
based on the Chern-Simons approach.
Our main results are presented in Sec.~\ref{sec: main_class_boson}.
We focus on SPT phases protected by $Z_N\rtimes R$ symmetry and
discuss the AKLT state on the square lattice.
Classification of 2D bosonic SPT phases under other related symmetries
[$Z_N\rtimes R$, U(1)$\rtimes R$, and U(1)$\times R$] is discussed
in Appendix.

\section{A brief review of the Chern-Simons approach}\label{sec: framework}
In this section, we briefly review the classification scheme of
2D bosonic SPT phases using the Chern-Simons approach.\cite{Lu_CS_2011}
In this scheme, SPT phases are classified
through stability analysis of gapless edge states.
The group structure of SPT phases is obtained
by studying the equivalence class of a direct sum of two phases.

\subsection{Classification scheme} 
In this paper we consider a class of 2D bosonic
SPT phases with nonchiral gapless edge modes.
We assume that SPT phases have $N_0$ pairs of
helical edge states and that
their bulk low-energy effective theory
is given by the Chern-Simons action
\begin{subequations}
\label{S_CS}
\begin{eqnarray}
S_\mathrm{CS}&=& \int\! d^2\bm{x} dt \mathcal{L}^0_{\mathrm{bulk}}, \\
\mathcal{L}^0_{\mathrm{bulk}}&=&
\frac{\epsilon_{\mu\nu\rho}}{4\pi}K_{I,J} 
a_{I\mu}(t,\bm{x})\partial_\nu a_{J\rho}(t,\bm{x}),
\end{eqnarray}
\end{subequations}
where $\epsilon_{\mu\nu\rho}$
is the totally anti-symmetric Levi-Civita tensor,
$(\partial_0,\partial_1,\partial_2)
=(\partial_t,\partial_{x_1},\partial_{x_2})$,
$\bm{x}=(x_1,x_2)$, 
and summation is assumed over repeated indices
$\mu,\nu,\rho=0,1,2$ and
$I,J=1,\ldots,2N_0$
(we adopt this convention throughout this paper).
The Chern-Simons gauge fields
$a_{I\mu}$ ($I=1,\cdots, 2N_0$)
describe low-energy dynamics in the gapped phase,
and $K$ is a symmetric matrix in $GL(2N_0,\mathbb{Z})$
with $\mathrm{det}K=(-1)^{N_0}$.
In bosonic systems every diagonal element of $K$-matrices is
an even integer.

The gapless boundary modes along the boundary (say, at $x_2=0$)
of the SPT phases are described by the boundary action
\begin{subequations}
\label{eq: S_edge}
\begin{eqnarray}
S_{\mathrm{edge}} &=&
\int dt dx_1 \mathcal{L}^0_\mathrm{edge},
\\
\mathcal{L}^0_\mathrm{edge} &=&
\frac{1}{4\pi}\left[
K_{I,J}(\partial_t \phi_I) (\partial_{x_1} \phi_J)
 - V_{I,J}(\partial_{x_1} \phi_I) (\partial_{x_1} \phi_J)
\right], \nonumber \\
&& \label{eq: L^0_edge}
\end{eqnarray}
\end{subequations}
where
$\phi_I$ $(I=1,\ldots,2N_0)$ are scalar bosonic fields
satisfying the equal-time commutation relation
\begin{equation}
[\phi_I(t,x_1),\partial_{x'_1} \phi_J(t,x'_1)]=
2\pi i (K^{-1})_{I,J}\delta(x_1-x'_1). 
\end{equation}
The matrix $V_{I,J}$ in Eq.\ (\ref{eq: L^0_edge}) is a nonuniversal
positive definite matrix. 
The operator $\partial_{x_1}\phi_I$ and the vertex operator
$:\!\!e^{i\phi_I}\!\!:$ are the density and the creation operators
of excitations in the $I$th edge mode.
Here colons denote normal ordering.

We assume that the boundary action $S_\mathrm{edge}$
and the bulk action $S_\mathrm{CS}$ are invariant under
a symmetry group $G$.
For the boundary action this means that
\begin{equation}
\mathcal{G}S_\mathrm{edge}\mathcal{G}^{-1}=S_\mathrm{edge}
\label{G S_edge G^-1 = S_edge}
\end{equation}
for any $\mathcal{G}\in G$
which induces linear transformation of the bosonic fields $\phi_I$
\begin{equation}
\mathcal{G} \bm{\phi} \mathcal{G}^{-1}
= U_\mathcal{G}\bm{\phi}
 +\delta \bm{\phi}_\mathcal{G},
\label{eq: trans_law general_G}
\end{equation}
where $\delta \bm{\phi}_\mathcal{G}$ is a constant $2N_0$-dimensional
vector with
$(\delta \bm{\phi}_\mathcal{G})_I \in [0,2\pi)$ ($I=1,\ldots, 2N_0$)
and the matrix $U_\mathcal{G}\in GL(2N_0,\mathbb{Z})$.
The transformation (\ref{eq: trans_law general_G})
for reflection and other local transformations will
be discussed in the following section.

A ground state with gapped excitations in the bulk is in an SPT phase
protected by a symmetry group $G$,
if it has gapless edge states that cannot be gapped
without symmetry breaking
by any perturbation allowed by the symmetry $G$.
Thus 2D SPT phases are classified by examining
the stability of their edge states against
perturbations of the form
\begin{equation}
 \mathcal{L}^\mathrm{int}_\mathrm{edge}
= \sum_{j=1}^{N_0}
 C_j :\!\cos({\bm{l}_j}\cdot\bm{\phi}+\alpha_j)\!:\, ,
\label{S^int_edge} 
\end{equation}
where $C_j$ and $\alpha_j$ are real constants,
and $\{\bm{l}_1,\ldots,\bm{l}_{N_0}\}$ is a set of linearly independent
vectors from $\mathbb{Z}^{2N_0}$.
The cosine terms in Eq.\ (\ref{S^int_edge}) are normal-ordered
as indicated by the colons.
However, we will omit colons for normal-ordered vertex
operators in the rest of this paper to simplify the notation.

The perturbations in $\mathcal{L}^\mathrm{int}_\mathrm{edge}$ are assumed
to fulfill the following conditions.
First, any pair of vectors, $(\bm{l}_j,\bm{l}_k)$, 
from the set $\{\bm{l}_1,\ldots,\bm{l}_{N_0}\}$
satisfies the Haldane's null vector condition,\cite{Haldane_Klein_1995}
\begin{equation}
\bm{l}^T_j K^{-1} \bm{l}_k = 0
\qquad(j,k=1,\ldots,N_0),
\label{eq: nullvec_cond_orignal}
\end{equation}
so that the linearly independent combinations of the bosonic fields
$\bm{l}_j\cdot\bm{\phi}$ ($j=1,\ldots,N_0$)
can be simultaneously pinned by the cosine potentials
in Eq.\ (\ref{S^int_edge}). 

Second, $\mathcal{L}^\mathrm{int}_\mathrm{edge}$ must be invariant
under the symmetry group $G$,
\begin{equation}
\mathcal{G}\mathcal{L}^\mathrm{int}_\mathrm{edge}\mathcal{G}^{-1}
=\mathcal{L}^\mathrm{int}_\mathrm{edge}.
\end{equation}
Since
$\mathcal{G}(\bm{l}_j\cdot\bm{\phi})\mathcal{G}^{-1}=
\bm{l}_j\cdot(U_{\mathcal{G}}\bm{\phi}+\delta\bm{\phi}_{\mathcal{G}})$,
the invariance of $\mathcal{L}^\mathrm{int}_\mathrm{edge}$
imposes the condition
\begin{equation}
\bm{l}_j^T K^{-1} U^T_{\mathcal{G}} \bm{l}_k = 0
\qquad(j,k=1,\ldots,N_0).
\label{eq: nullvec_cond_symm}
\end{equation}

The third condition is concerned with the absence of spontaneous
symmetry breaking.
The symmetry $G$ can be spontaneously broken in the ground state
when $N_0$ linearly independent fields $\bm{l}_j\cdot\bm{\phi}$
($j=1,\ldots,N_0$) are pinned,
even if the interactions in $\mathcal{L}^\mathrm{int}_\mathrm{edge}$
respect the symmetry.
To see this, let us define from the vectors
$\{\bm{l}_1,\cdots,\bm{l}_{N_0}\}$
a set of vectors
\begin{equation}
L:=\left\{\bm{l}\Big|\,
\bm{l}=\sum^{N_0}_{n=1}j_n\bm{l}_n,
~j_n\in\mathbb{Z}~(n=1,\cdots,N_0)\right\}.
\end{equation}
Any field $\bm{l}\cdot\bm{\phi}$ ($\bm{l}\in L$)
takes a constant expectation value in the ground state.
We then define from $L$ another set of vectors
\begin{equation}
\widetilde{L}:=
\left\{\tilde{\bm{l}}\,
\Big|
\,\tilde{\bm{l}}=
\frac{\bm{l}}{\mathrm{gcd}(l_1,\cdots,l_{2N_0})},
~\bm{l}=(l_1,\cdots,l_{2N_0})\in L\right\},
\label{tilde l}
\end{equation}
where gcd denotes the greatest common divisor of the integers
in the parentheses.
Note that $\widetilde{L}\supseteq L$.
The set $\widetilde{L}$ is a Bravais lattice,
whose primitive lattice vectors are denoted by
$\bm{v}_1,\cdots,\bm{v}_{N_0}$.
The elementary bosonic fields
\begin{equation}
\bm{v}_n\cdot\bm{\phi} \quad (n=1,\ldots,N_0)
\label{eq: v alpha}
\end{equation}
take constant expectation values in the ground state.
If the expectation values are invariant,
i.e.,
\begin{equation}
\langle\bm{v}_n\cdot(U_{\mathcal{G}}\bm{\phi}
+\delta\bm{\phi}_{\mathcal{G}})\rangle
=\langle\bm{v}_n\cdot\bm{\phi}\rangle
\qquad(n=1,\ldots,N_0)
\end{equation}
modulo $2\pi$
for any $\mathcal{G}\in G$,
then the edge modes can be gapped without any symmetry breaking.
Otherwise, the ground state breaks the symmetry $G$ spontaneously.

If there exists a set of vectors $\{\bm{l}_1,\ldots,\bm{l}_{N_0}\}$
that satisfies the above three conditions,
then the edge states
can be gapped out without symmetry breaking
[with strong enough $C_j$ even when
$\cos(\bm{l}_j\cdot\bm{\phi})$ is an irrelevant operator in
the renormalization-group sense],
and the resulting gapped phase is a trivial phase.
On the other hand, if we cannot find such a set of vectors
$\{\bm{l}_1,\ldots,\bm{l}_{N_0}\}$, then the edge states are stable,
and the bulk ground state realizes a 2D SPT phase.

SPT phases form an Abelian group as follows.\cite{Lu_CS_2011}
Elements of the Abelian group are equivalence classes
of phases that are connected without a closing of a gap
under adiabatic deformation of the
action (Hamiltonian) while preserving the symmetry $G$.
The SPT phases described by the action (\ref{eq: S_edge}) and
the transformation (\ref{eq: trans_law general_G})
under the symmetry group $G$
are denoted by
$\Psi_G[K, \{U_{\mathcal{G}},\delta \bm{\phi}_{\mathcal{G}}\} ]$,
while a trivial phase is denoted by ``$0$".
The summation of two phases is defined as a direct sum of
the two phases,
\begin{align}
&\Psi_G[K, \{U_\mathcal{G},\delta \bm{\phi}_\mathcal{G}\} ] 
\oplus \Psi_G[K', \{U'_\mathcal{G},\delta \bm{\phi}'_\mathcal{G}\} ] \n
&\qquad
=\Psi_G[K\oplus K', \{U_\mathcal{G}\oplus U'_\mathcal{G},
        \delta \bm{\phi}_\mathcal{G} \oplus \delta \bm{\phi}'_\mathcal{G}\} ].
\end{align}
The bosonic fields in the direct sum of two phases,
$\bm{\phi}=(\phi_1,\ldots,\phi_{4N_0})^T$, are transformed
by $\mathcal{G}\in G$ as
\begin{equation}
\mathcal{G} \bm{\phi} \mathcal{G}^{-1}
= (U_\mathcal{G}\oplus U'_\mathcal{G}) \bm{\phi}
 + \delta \bm{\phi}_\mathcal{G}\oplus\delta \bm{\phi}'_\mathcal{G}.
\end{equation}
with $\mathcal{G}\in G$.
The inverse element of an SPT phase
$\Psi_G[K,\{U_\mathcal{G},\delta\bm{\phi}_\mathcal{G}\}]$ is
found from the relation
\begin{align}
\Psi_G[-K, \{U_\mathcal{G},\delta \bm{\phi}_\mathcal{G}\} ]
\oplus \Psi_G[K, \{U_\mathcal{G},\delta \bm{\phi}_\mathcal{G}\} ]
=0,
\label{eq: inverse element}
\end{align}
which is understood by noting that the fields
$(\phi_1-\phi_{2N_0+1},\phi_2-\phi_{2N_0+2},\ldots,\phi_{2N_0}-\phi_{4N_0})$
can be pinned without symmetry breaking by the pinning potential
\begin{equation}
\mathcal{L}^\mathrm{int}_\mathrm{edge}
=\sum^{2N_0}_{j=1}C_j\cos(\phi_j-\phi_{2N_0+j}).
\end{equation}
Thus, the equivalence relation between
$\Psi_G[K, \{U_\mathcal{G},\delta \bm{\phi}_\mathcal{G}\} ]$ and
$\Psi_G[K, \{U'_\mathcal{G},\delta \bm{\phi}'_\mathcal{G}\} ]$ is
identical to the equivalence relation between
$\Psi_G[K, \{U_\mathcal{G},\delta \bm{\phi}_\mathcal{G}\} ]\oplus
 \Psi_G[-K, \{U'_\mathcal{G},\delta \bm{\phi}'_\mathcal{G}\} ]$
and a trivial phase ``0''.

The addition rule of SPT phases allows us to construct SPT phases
of a larger number of degrees of freedom from small building blocks.
Following Lu and Vishwanath,\cite{Lu_CS_2011}
we consider a minimal model of SPT phases with
a pair of helical edge states, described by
the edge theory $S_\mathrm{edge}$ of
two bosonic fields
$\phi_I$ ($I=1,2$) with a $2\times2$ $K$-matrix
of $\mathrm{det}K=-1$.
SPT phases with multiple pairs of helical edge states
are obtained by combining minimal SPT phases,
i.e., taking a direct sum of minimal models.
Thus, we take
\begin{equation}
K=\sigma^x,
\end{equation}
unless otherwise noted,
since $2\times2$ $K$-matrices for bosonic systems can be
reduced to $K=\sigma^x$. \cite{Lu_CS_2011}

Finally we note that there is a redundancy in the representation
in Eq.~{(\ref{eq: trans_law general_G})}. 
The action $S^0_\mathrm{edge}=\int dx_1 dt \mathcal{L}^{0}_\mathrm{edge}$ is
unchanged by substituting the fields
$\bm{\phi}' = X \bm{\phi} +\Delta \bm{\phi}$ with
$\Delta \bm{\phi} \in \mathbb{R}^{2}$ and
$X \in GL(2,\mathbb{Z})$ satisfying $X^T K X=K$.
Two representations
$\{ U_\mathcal{G}, \delta \bm{\phi}_\mathcal{G}\}$ and
$\{ U'_\mathcal{G}, \delta \bm{\phi}'_\mathcal{G}\}$ are therefore equivalent
if they satisfy
\begin{eqnarray}
\delta \bm{\phi}'_\mathcal{G}
&=& X [\delta \bm{\phi}_\mathcal{G}
 + (\1-U_\mathcal{G})X^{-1}\Delta \bm{\phi} ],
\nonumber \\
U'_\mathcal{G} &=& X U_\mathcal{G} X^{-1}.
\label{eq: scheme_trans_gauge}
\end{eqnarray}

\section{SPT phases protected by reflection symmetry and $Z_N$ symmetry} \label{sec: main_class_boson}
\begin{figure}[!h]
\begin{center}
\includegraphics[width=8cm,clip]{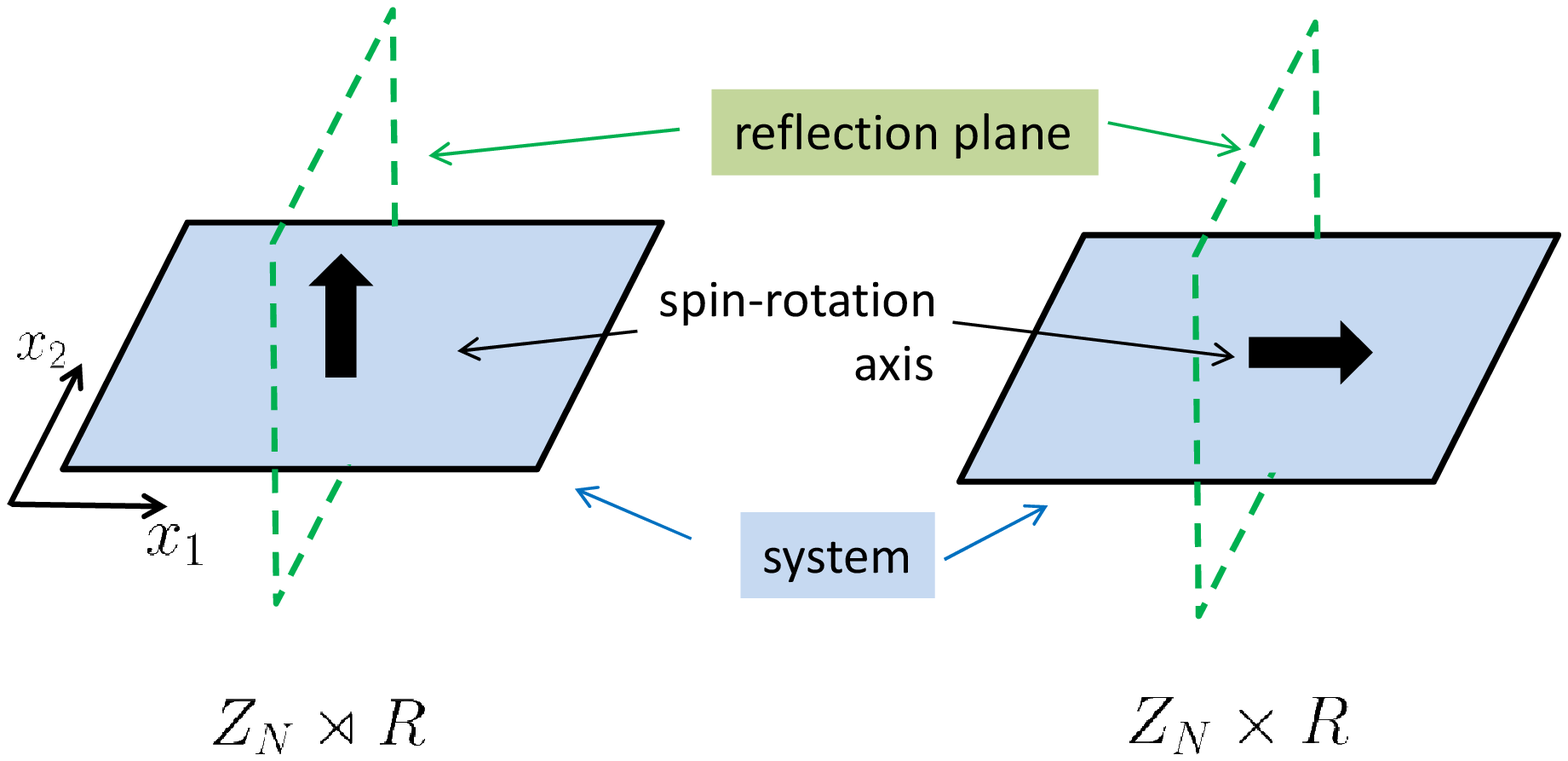}
\end{center}
\caption{
(Color online) Schematic picture of 2D systems 
with reflection symmetry and spin-rotation symmetry.
The reflection planes (dashed green rectangles) are perpendicular to
the 2D systems (blue planes).
The spin-rotation axes are represented by black arrows. 
The symmetry group is $Z_N \rtimes R$ when the rotation axis is
parallel to the reflection plane (left panel),
while the symmetry group is $Z_N \times R$ when the rotation axis
is perpendicular to the reflection plane (right panel).
}
\label{fig: rot_axis}
\end{figure}

In this section, we classify bosonic SPT phases protected by
reflection symmetry $R$ and discrete local symmetry $Z_N$.
There are two possible group structures for these symmetries:
(i) $Z_N \rtimes R$ and (ii) $Z_N \times R$.
When the $Z_N$ symmetry corresponds to $2\pi/N$ rotation
in integer-spin systems, these two group structures are realized
(a) when the spin-rotation axis is parallel to the reflection plane and
(b) when the spin-rotation axis is perpendicular to the reflection plane,
respectively (see Fig.\ \ref{fig: rot_axis}). 
In this section we focus on the $Z_N \rtimes R$ symmetry
and show that an example of the SPT phases protected by this symmetry
is given by the $S=2$ Affleck-Kennedy-Lieb-Tasaki (AKLT) state
on the square lattice.
The classification of SPT phases protected by (ii) $Z_N \times R$ symmetry,
(iii) U(1)$\rtimes R$ symmetry, and (iv) U(1)$\times R$ symmetry
is discussed in Appendix.
The results of the classification are summarized in
Table \ref{table: Bose_table}.

\begin{table*}[htb]
\begin{center}
\caption{
Classification of 2D bosonic SPT phases protected by reflection symmetry $R$
and additional local symmetry [$Z_N$ or U(1)].
For each symmetry group denoted in the first column and the parity of $N$
of the $Z_N$ symmetry in the second column,
the Abelian groups of SPT phases are listed in the third column.
The Abelian group of SPT phases protected by
$\mathrm{U(1)} \rtimes R$ and $\mathrm{U(1)} \rtimes R$ are
in agreement with the results summarized in Table III
of Ref.~\onlinecite{Hsieh_CS_CPT_2014}.}
\begin{center}
\begin{tabular}{cccc} \hline \hline 
~Symmetry~ & ~Parity of $N$~ & ~Abelian group of SPT phases~ \\ \hline
$Z_N \rtimes R$  & Even & $\mathbb{Z}_2\times\mathbb{Z}_2$  \\ 
                 & Odd  & $0$                 \\
$Z_N \times R$   & Even & $\mathbb{Z}_2\times\mathbb{Z}_2$  \\
                 & Odd  & $0$                \\ 
$\mathrm{U(1)} \rtimes R$  & --  & $\mathbb{Z}_2$  \\ 
$\mathrm{U(1)} \times R$   & --  & $0$             \\
\hline \hline
\end{tabular}
\end{center}
\label{table: Bose_table}
\end{center}
\end{table*}

\subsection{$Z_N \rtimes R$} \label{subsection: boson_zn_rt_R} 
In this subsection, we apply the Chern-Simons approach
to 2D bosonic SPT phases protected by the symmetry group $Z_N\rtimes R$.

First, we determine the transformation laws
[Eq.~(\ref{eq: trans_law general_G})]
of bosonic fields $\bm{\phi}_I$'s under $R$ and $Z_N$.
The invariance of the Lagrangian $\int\!dx_1\mathcal{L}^0_\mathrm{edge}$
under the reflection $R$,
\begin{eqnarray}
&&
\int\! dx_1 (U^T_R K U_R)_{I,J}
\partial_{t} \phi_I (t,-x_1) \partial_{x_1} \phi_J(t,-x_1) \nonumber \\
&&=
\int\! dx_1 K_{I,J} \partial_{t} \phi_I (t,x_1) \partial_{x_1} \phi_J(t,x_1),
\end{eqnarray}
is guaranteed if $\phi_I$'s obey the transformation
\begin{equation}
R\, \bm{\phi} (t,x_1)R^{-1}
= U_R\bm{\phi}_J(t,-x_1)+\delta \bm{\phi}_{R},
\label{R transformation_1}
\end{equation}
where the matrix $U_R \in GL(2,\mathbb{Z})$ satisfies
\begin{equation}
U^T_R K U_R = -K.
\label{R transformation_2}
\end{equation}
This condition is satisfied by $U_R=\pm\sigma^z$ for $K=\sigma^x$.
Since the two representations $U_R=\sigma^z$ and $U_R=-\sigma^z$ are
related by the transformation 
with $X=\sigma^x$
[Eq.\ (\ref{eq: scheme_trans_gauge})],
it suffices to take the representation
\begin{equation}
U_R=-\sigma^z.
\end{equation}

Similarly, the Lagrangian $\mathcal{L}^0_\mathrm{edge}$ is invariant
if the bosonic fields $\phi_I$'s are transformed
by $g\in Z_N$ as
\begin{equation}
g \bm{\phi}(x) {g}^{-1} = U_{g} \bm{\phi}(x) +\delta \bm{\phi}_{g}
\label{g transformation_1}
\end{equation}
with
a matrix $U_{g} \in GL(2,\mathbb{Z})$ satisfying
\begin{equation}
U^T_{g} K U_{g} = K.
\label{g transformation_2}
\end{equation}
Any choice from $U_g=\pm \1, \pm\sigma^x$ fulfills this condition.
However, we discard the representations $U_g=\pm \sigma^x=\pm K$,
since they are not compatible with Eq.\ (\ref{eq: nullvec_cond_symm}) for $j=k$.
Here we take the representation
\begin{equation}
U_g=\1,
\end{equation}
since it is realized in spin models in which
the $Z_N$ symmetry corresponds to spin-rotation symmetry
(see Sec.~\ref{subsec: AKLT}).
We do not consider the other case $U_g=-\1$ in this paper.
Incidentally, this case with $N=2$ can be relevant to bosonic systems
with charge conjugation symmetry,
where creation operators of quasiparticles $:\!e^{i\phi_I}\!\!:$ are
transformed to annihilation operators.

The representation $\{U_\mathcal{G}, \delta \bm{\phi}_\mathcal{G} \}$ of
symmetry operation $\mathcal{G}\in G$ is constrained by the group structure
of the symmetry group $G$.
For the symmetry group $G=Z_N \rtimes R$,
the generators of the group satisfy the relation $R^2=e=g^N$,
where $e$ denotes the identity element of $G$.
Accordingly, the representation
$\{U_\mathcal{G},\delta\bm{\phi}_\mathcal{G}\}$
must satisfy the conditions
\begin{eqnarray}
U^2_R \bm{\phi} +(\1+U_R)\delta \bm{\phi}_R
&=&\bm{\phi},
\label{eq: R2=e}
\\
U^N_g \bm{\phi} + \sum^{N-1}_{k=0} U^k_g \delta \bm{\phi}_g
&=&\bm{\phi},
\label{eq: gN=e}
\end{eqnarray}
where $\1$ is a $2 \times 2$ unit matrix.
Furthermore, the algebraic relation $RgRg=e$ obeyed by
the generators of the symmetry group $G=Z_N \rtimes R$
leads to the additional condition
\begin{equation}
U_{g} U_R U_{g} U_R \bm{\phi}
+ (\1+ U_gU_R) (U_g \delta \bm{\phi}_R +\delta \bm{\phi}_g)
= \bm{\phi}. 
\label{eq: RgRg=e}
\end{equation}

In the following we discuss cases where $N$ is even and odd separately.

\subsubsection{Even $N$}\label{subsubsection: boson_zn_rt_R even}
Given the representation
\begin{equation}
U_R=-\sigma^z,
\qquad
U_g=\1,
\end{equation}
we deduce from Eqs.~(\ref{eq: R2=e})--(\ref{eq: RgRg=e})
the transformation laws for bosonic fields
\begin{subequations}
\label{eq: g_rt_R trans_law Ug=1}
\begin{eqnarray}
g \bm{\phi} g^{-1}&=& \bm{\phi} + \frac{2\pi}{N}
\begin{pmatrix}
k_g \\ 0
\end{pmatrix}
+ \pi 
\begin{pmatrix}
0 \\ n_g
\end{pmatrix}
,
\label{g phi g^-1 even N}  \\
R \bm{\phi} R^{-1}&=& -\sigma^z \bm{\phi} + \pi  
\begin{pmatrix}
0 \\n_R
\end{pmatrix}
,
\label{R phi R^-1}
\end{eqnarray}
where
\begin{eqnarray}
n_g,n_R=0,1, \;\;\; k_g=0,\ldots,N-1. 
\end{eqnarray}
\end{subequations}
In Eq.\ (\ref{R phi R^-1})
the phase shift $\delta\phi_1$ caused by the reflection $R$ is
set equal to zero by the basis transformation
in Eq.~{(\ref{eq: scheme_trans_gauge})} with $X=\1$
and $\Delta\phi_1$ chosen appropriately.

Let us label by a set of integers $[k_g,n_g,n_R]$ a topological phase
in which bosonic fields are transformed as in
Eqs.\ (\ref{eq: g_rt_R trans_law Ug=1}).
We will show that the SPT phases form an Abelian group
$\mathbb{Z}_2\times\mathbb{Z}_2$
by proving the following three properties:
\begin{itemize}
 \item[(a)] Phases $[0,n_g,n_R]$ and $[k_g,0,0]$ are trivial
 ($[0,n_g,n_R]=[k_g,0,0]=0$).
 \item[(b)] Any phase is generated from $[1,0,1]$ and $[1,1,0]$,
 which satisfy $[1,0,1]\oplus[1,0,1]=[1,1,0]\oplus[1,1,0]=0$.
 \item[(c)] The two phases $[1,0,1]$ and $[1,1,0]$ are
 independent generators of SPT phases.
\end{itemize}

\textit{Proof of (a):}
The null vector condition [Eq.~{(\ref{eq: nullvec_cond_orignal})}]
with $K=\sigma^x$ allows only pinning potentials of the form
$\cos (l\phi_1+\alpha_l)$ or $\cos (l\phi_2+\alpha_l)$
with $l \in \mathbb{Z}$ and $\alpha_l\in \mathbb{R}$.
When $k_g=0$, the pinning potential
\begin{equation}
H_{\mathrm{int}}= C \int\! dx_1 \cos(\phi_1),
\end{equation}
is invariant under the transformations
in Eq.~(\ref{eq: g_rt_R trans_law Ug=1})
and can pin the field $\phi_1$ at $\langle \phi_1 \rangle=0$ or $\pi$
depending on the sign of $C$.
No symmetry is broken by the pinning.
Thus, the phase $[0,n_g,n_R]$ is reduced to a trivial insulator.
When $n_g=n_R=0$, the pinning potential
\begin{equation}
H_{\mathrm{int}}=
C \int\! dx_1 \cos(\phi_2+\alpha),
\end{equation}
is invariant under the transformations
in Eq.~(\ref{eq: g_rt_R trans_law Ug=1})
and can pin the field $\phi_2$ at $\langle \phi_2+\alpha\rangle=0$ or $\pi$
without symmetry breaking.
Thus, the phase $[k_g,0,0]$ is a trivial insulator.

\textit{Proof of (b):}
We first show the following addition relations
of SPT phases:
\begin{subequations}
\label{eq: g_rt_R Ug=1 sum}
\begin{eqnarray}
[k_g,n_g,n_R] \oplus [k_g,n'_g,n'_R] &=& [k_g,n_g+n'_g,n_R+n'_R], \nonumber\\
&&\\
{}[k_g,n_g,n_R] \oplus [k_g',n_g,n_R] &=& [k_g+k_g',n_g,n_R].
\end{eqnarray}
\end{subequations}
The composition of two phases $[k_g,n_g,n_R]$ and $[k_g,n'_g,n'_R]$
has bosonic fields $\bm{\phi}=(\phi_1,\phi_2,\phi_3,\phi_4)^T$
and a $K$-matrix $K=\sigma^x\oplus\sigma^x$.
The fields obey the commutation relations
\begin{equation}
[\phi_I(x_1),\partial_{x'_1}\phi_J(x'_1)]=
2\pi i (\sigma^x\oplus\sigma^x)_{I,J}\delta(x_1-x'_1),
\end{equation}
and the transformation laws
\begin{subequations}
\begin{eqnarray}
{}g \bm{\phi} g^{-1} &=&
\bm{\phi} + \frac{2\pi k_g}{N}(\bm{e}_1+\bm{e}_3)
 + \pi n_g \bm{e}_2 + \pi n'_g \bm{e}_4, \nonumber \\
 && \\
R \bm{\phi} R^{-1} &=&
-(\sigma^z\oplus\sigma^z) \bm{\phi} + \pi n_R \bm{e}_2 +\pi n'_R \bm{e}_4,
\end{eqnarray}
with
\begin{align}
k_g&=0,1,\ldots, N-1,  &
n_g,n'_g,n_R,n'_R&=0,1.
\end{align}
\end{subequations}
Here, $\bm{e}_j$ ($j=1,\ldots, 4$) denotes the $j$th unit vector,
$(e_{j})_I=\delta_{j,I}$.
We now make a basis transformation and define a new set of bosonic fields
\begin{equation}
\bm{\psi}=
(\psi_1,\psi_2,\psi_3,\psi_4)^T
=(\phi_1-\phi_3,\phi_2,\phi_3,\phi_2+\phi_4)^T,
\end{equation}
which have the same $K$-matrix and commutators
\begin{equation}
[\psi_I(x),\partial_{x'} \psi_J(x')]=
2\pi i (\sigma^x \oplus \sigma^x)_{I,J} \delta (x-x').
\end{equation}
Without pinning potentials, there are two pairs of gapless helical edge
modes: $(\psi_1,\psi_2)$ and $(\psi_3,\psi_4)$.
A potential of the form
\begin{equation}
H_{\mathrm{int}}= C \int\! dx_1 \cos(\psi_1)
\end{equation}
can pin the $\psi_1$ field and gap out the $(\psi_1,\psi_2)$ sector
without symmetry breaking.
The helical edge states in the $(\psi_3,\psi_4)$ sector remains gapless
and correspond to the phase $[k_g,n_g+n_g',n_R+n_R']$.
Equation (\ref{eq: g_rt_R Ug=1 sum}a) follows.

In a similar way, we obtain Eq.\ (\ref{eq: g_rt_R Ug=1 sum}b)
by making basis transformation
\begin{equation}
\bm{\psi}'=
(\psi_1',\psi_2',\psi_3',\psi_4')^T
=(\phi_1+\phi_3,\phi_2,\phi_3,\phi_2-\phi_4)^T
\end{equation}
and adding a potential of the form
\begin{equation}
H_{\mathrm{int}}= C\int\!dx_1 \cos(\psi'_4+\alpha).
\end{equation}
In this case the $(\psi'_3,\psi'_4)$ sector is a trivial gapped state
and can be discarded.
The edge states in the remaining $(\psi'_1,\psi'_2)$ sector
corresponds to the phase $[k_g+k_g',n_g,n_R]$, 
and thus we obtain Eq.\ (\ref{eq: g_rt_R Ug=1 sum}b).

We find from Eqs.\ (\ref{eq: g_rt_R Ug=1 sum}) that
\begin{subequations}
\begin{align}
[1,1,0]\oplus[1,1,0] &=[1,2,0]=[1,0,0]=0, \\
[1,0,1]\oplus[1,0,1] &=[1,0,2]=[1,0,0]=0,
\end{align}
\end{subequations}
since phase shifts are defined modulo $2\pi$.
Furthermore, 
using Eqs.\ (\ref{eq: g_rt_R Ug=1 sum}) successively,
we can reduce any phase $[k_g,n_g,n_R]$
to four phases:
\begin{equation}
[k_g,n_g,n_R]=
\left\{
\begin{array}{ll}
0, & (k_gn_g,k_gn_R)=(\mbox{e},\mbox{e}),\\
\mbox{[1,1,0]}, & (k_gn_g,k_gn_R)=(\mbox{o},\mbox{e}),\\
\mbox{[1,0,1]}, & (k_gn_g,k_gn_R)=(\mbox{e},\mbox{o}),\\
\mbox{[1,1,0]}\oplus\mbox{[1,0,1]}, & (k_gn_g,k_gn_R)=(\mbox{o},\mbox{o}),
\end{array}
\right.
\label{four cases}
\end{equation}
where ``e'' and ``o'' stand for ``even'' and ``odd'', respectively.

\textit{Proof of (c):} 
We show that the two phases $[1,0,1]$ and $[1,1,0]$ are neither
equivalent to each other nor connected to the trivial phase 0.
To this end, we show that edge modes of the phase
$[\bar{k}_g,n_g,n_R]\oplus[\bar{k}'_g,n'_g,n'_R]^{-1}$
with $\bar{k}_g,\bar{k}'_g=0,1$ cannot be gapped out,
unless $(\bar{k}_g,n_g,n_R)=(\bar{k}'_g,n'_g,n'_R)$
or $(\bar{k}_g,0,0)=(0,n'_g,n'_R)$.
It follows from Eq.~(\ref{eq: inverse element}) that
the phase $[\bar{k}_g,n_g,n_R]\oplus[\bar{k}'_g,n'_g,n'_R]^{-1}$
has edge modes described by
the bosonic fields $\bm{\phi}=(\phi_1,\phi_2,\phi_3,\phi_4)^T$
with a $K$-matrix $K=\sigma^x\oplus (-\sigma^x)$.
The bosonic fields
$(\phi_1,\phi_2)$ and $(\phi_3,\phi_4)$ obey
the transformation laws of $[\bar{k}_g,n_g,n_R]$ and
$[\bar{k}'_g,n'_g,n'_R]^{-1}$, respectively.

Gapping the bosonic fields
$\bm{\phi}=(\phi_1,\phi_2,\phi_3,\phi_4)^T$ requires
two pinning potentials $\cos (\bm{l}_1 \cdot \bm{\phi}+\alpha_1)$
and $\cos (\bm{l}_2 \cdot \bm{\phi}+\alpha_2)$,
whose integer vectors $\bm{l}_1$ and $\bm{l}_2$ must satisfy
Eqs.~(\ref{eq: nullvec_cond_orignal}) and (\ref{eq: nullvec_cond_symm}),
or equivalently,
\begin{subequations}
\label{eq: R-inv_null_vec_cond}
\begin{eqnarray}
\bm{l}^T_i[\sigma^x\oplus(-\sigma^x)]\bm{l}_j&=&0,\\
\bm{l}^T_i[i\sigma^y\oplus(-i\sigma^y)]\bm{l}_j&=&0,
\end{eqnarray}
\end{subequations}
for $i,j=1,2$.
Solutions to these equations are given by
\begin{subequations}
\label{eq: R-inv_null_vectors}
\begin{eqnarray}
\bm{l}_1&=& (\alpha p,\beta q,\alpha q, \beta p)^T, \\
\bm{l}_2&=& (\alpha' p,\beta' q,\alpha' q, \beta' p)^T.
\end{eqnarray}
\end{subequations}
with $\alpha,\beta,\alpha',\beta',p,q\in \mathbb{Z}$. %
\footnote{
This is obtained as follows.
The generic solution to the equation $\bm{l}^TK^{-1}\bm{l}=0$
with $K=\sigma^x\oplus(-\sigma^x)$ is given by
\begin{eqnarray*}
\bm{l}&=&(pm,qn,qm,pn)^T\\
&=&
p(m,0,0,n)+q(0,n,m,0)\\
&=&
m(p,0,q,0)+n(0,q,0,p),
\end{eqnarray*}
where $n,m,p,q\in \mathbb{Z}$.
Thus the two linearly independent vectors $\bm{l}_1$ and $\bm{l}_2$
satisfying Eq.~(\ref{eq: R-inv_null_vec_cond}a) have the form
\begin{eqnarray*}
\alpha\bm{a}+\beta\bm{b}, &\quad& \alpha'\bm{a}+\beta'\bm{b},
\end{eqnarray*}
where $\alpha$, $\beta$, $\alpha'$, and $\beta'$ are integers
with $\alpha\beta'-\beta\alpha'\ne0$.
Here we have two possible choices for the vectors $\bm{a}$ and $\bm{b}$:
\[
\bm{a}=(m,0,0,n)^T, \quad \bm{b}=(0,n,m,0)^T,
\]
or
\begin{eqnarray*}
\bm{a}=(p,0,q,0)^T, &\quad& \bm{b}=(0,q,0,p)^T.
\end{eqnarray*}
It is straightforward to check that the latter set of vectors
is compatible with the condition posed by the reflection symmetry
[Eq.~(\ref{eq: R-inv_null_vec_cond}b)],
while the former is not.
}
If $pq=0$ and $p\ne q$, then the elementary bosonic fields
defined in Eq.~(\ref{eq: v alpha})
are given by
$(\bm{v}_1\cdot\bm{\phi},\bm{v}_2\cdot\bm{\phi})=(\phi_3,\phi_2)$
or $(\phi_1,\phi_4)$.
If $pq\ne0$, we can assume $\mathrm{gcd}(p,q)=1$
and obtain the elementary bosonic fields
$\bm{v}_1 \cdot\bm{\phi}= p\phi_1+q\phi_3$ and
$\bm{v}_2 \cdot\bm{\phi}= q\phi_2+p\phi_4$.
In either case the fields are transformed as
\begin{subequations}
\label{eq: g_rt_R va vb}
\begin{eqnarray}
g (\bm{v}_1 \cdot\bm{\phi}) g^{-1} &=&
 \bm{v}_1 \cdot\bm{\phi} + \frac{2\pi}{N}(p\bar{k}_g+q\bar{k}'_g),
\label{g v_1 g^-1} \\
R (\bm{v}_1 \cdot\bm{\phi}) R^{-1} &=& -\bm{v}_1 \cdot\bm{\phi},
\label{R v_1 R^-1} \\
g (\bm{v}_2 \cdot\bm{\phi}) g^{-1} &=&
 \bm{v}_2 \cdot\bm{\phi} + \pi(qn_g+pn'_g),
\label{g v_2 g^-1} \\
R (\bm{v}_2 \cdot\bm{\phi}) R^{-1} &=& \bm{v}_2 \cdot\bm{\phi}+\pi(qn_R+pn'_R),
\label{R v_2 R^-1}
\end{eqnarray}
\end{subequations}
where we assume $(p,q)=(1,0)$ or $(0,1)$ if $pq=0$.
When $(p,q)=(\mathrm{odd},\mathrm{odd})$, the phase shifts in
Eqs.\ (\ref{g v_1 g^-1}), (\ref{g v_2 g^-1}), and (\ref{R v_2 R^-1})
are equal to zero (mod $2\pi$) only when
$(\bar{k}_g,n_g,n_R)=(\bar{k}'_g,n'_g,n'_R)$.
This means that the edge modes cannot be gapped out without symmetry breaking
unless $(\bar{k}_g,n_g,n_R)=(\bar{k}'_g,n'_g,n'_R)$.
Similarly, when $(p,q)=(\mathrm{even},\mathrm{odd})$,
the edge modes can be gapped out without symmetry breaking
only if $n_g=n_R=\bar{k}_g'=0$, i.e.,
$[\bar{k}_g,n_g,n_R]=[\bar{k}_g',n_g',n_R']=0$.
Thus, the two phases $[1,0,1]$ and $[1,1,0]$ are inequivalent, and
both of them are distinct from the trivial phase.

From (a), (b), and (c),
we conclude that the Abelian group of the SPT phases
protected by $Z_N\rtimes R$ is $\mathbb{Z}_2\times\mathbb{Z}_2$
generated by $[1,0,1]$ and $[1,1,0]$.

Finally, we note that the SPT phase $[1,1,0]$ is stable
even in the absence of the reflection symmetry,
while the other two SPT phases, $[1,0,1]$ and $[1,1,0]\oplus[1,0,1]$,
are SPT phases that are stable only in the presence of
both the reflection symmetry and the $Z_N$ symmetry.

\subsubsection{Odd $N$} \label{subsubsection: boson_zn_rt_R odd}
When $N$ is odd, the bosonic fields $\bm{\phi}=(\phi_1,\phi_2)^T$
are transformed under symmetry operations as
\begin{subequations}
\label{eq: g_rt_R trans_law Ug=1 odd N}
\begin{eqnarray}
g \bm{\phi} g^{-1}&=& \bm{\phi} + \frac{2\pi}{N}
\begin{pmatrix}
k_g \\ 0
\end{pmatrix}
, 
\label{g phi g^-1 odd N} \\
R \bm{\phi} R^{-1}&=& -\sigma^z \bm{\phi} + \pi  
\begin{pmatrix}
0 \\n_R
\end{pmatrix}
,  
\end{eqnarray}
with
\begin{eqnarray}
n_R=0,1, \;\;\; k_g=0,\ldots,N-1. 
\end{eqnarray}
\end{subequations}
We note that the above transformation rules are obtained from
Eqs.\ (\ref{eq: g_rt_R trans_law Ug=1}) by setting $n_g=0$.
The vanishing phase shift of $\phi_2$ under the $g$ transformation
is a consequence of the conditions (\ref{eq: gN=e}) and
(\ref{eq: RgRg=e}) for odd $N$.

There is no SPT phase when $N$ is odd.
This conclusion is obtained by using the classification for even $N$
discussed above.
Let us label phases by a set of integers $[k_g,n_R]$.
Imposing $n_g=0$ in Eq.\ (\ref{four cases}), we find
\begin{eqnarray}
{}[k_g,n_R]&=&
\left\{
\begin{array}{ll}
\mbox{0}, & k_gn_R=\mbox{even},\\
\mbox{[1,1]}, & k_gn_R=\mbox{odd}.
\end{array}
\right.
\end{eqnarray}
Next we prove that two phases $[0,1]\,(=0)$ and $[1,1]$ are equivalent
by showing that the edge modes of the phase $[0,1]\oplus[1,1]^{-1}$
can be gapped without symmetry breaking (i.e., $[0,1]\oplus[1,1]^{-1}=0$).
The phase $[0,1]\oplus[1,1]^{-1}$ has edge modes described 
by bosonic fields $\bm{\phi}=(\phi_1,\phi_2,\phi_3,\phi_4)^T$
with a $K$-matrix $K=\sigma^x\oplus(-\sigma^x)$.
The bosonic fields
$(\phi_1,\phi_2)$ and $(\phi_3,\phi_4)$ 
obey the transformation laws of $[0,1]$ and $[1,1]$, respectively.
These fields are gapped by the pinning potential of the form
\begin{equation}
H_\mathrm{int}=C_1\int\! dx_1\cos(\bm{l}_1\cdot\bm{\phi})
               +C_2\int\! dx_1\cos(\bm{l}_2\cdot\bm{\phi}+\alpha), 
\label{eq: odd N potential}
\end{equation}
where
$\bm{l}_1\cdot\bm{\phi}=\phi_1+N\phi_3$ and
$\bm{l}_2\cdot\bm{\phi}=N\phi_2+\phi_4$.
These integer vectors
$\bm{l}_1$ and $\bm{l}_2$ satisfy the conditions
in Eq.\ (\ref{eq: R-inv_null_vec_cond}).
The corresponding elementary bosonic variables,
\begin{subequations}
\begin{eqnarray}
\bm{v}_1\cdot\bm{\phi}&=&\bm{l}_1\cdot\bm{\phi}=\phi_1+N\phi_3, \\
\bm{v}_2\cdot\bm{\phi} &=&\bm{l}_2\cdot\bm{\phi}=N\phi_2+\phi_4,
\end{eqnarray}
\end{subequations}
are transformed as
\begin{subequations}
\begin{eqnarray}
g (\bm{v}_1\cdot\bm{\phi}) g^{-1} &=& \bm{v}_1\cdot\bm{\phi} +2\pi, \\
R (\bm{v}_1\cdot\bm{\phi}) R^{-1} &=& -\bm{v}_1\cdot\bm{\phi}, \\
g (\bm{v}_2\cdot\bm{\phi}) g^{-1} &=& \bm{v}_2\cdot\bm{\phi}, \\
R (\bm{v}_2\cdot\bm{\phi}) R^{-1} &=& \bm{v}_2\cdot\bm{\phi}+(N+1)\pi. 
\end{eqnarray}
\end{subequations}
Since the phase shift $(N+1)\pi$ is a multiple of $2\pi$ for odd $N$,
the edge modes can be gapped out without symmetry breaking by
the pinning potential.
We thus obtain $[0,1]\oplus[1,1]^{-1}=0$, thereby $[1,1]=[0,1]=0$.
Hence we conclude that $[k_g,n_R]=0$ for any $k_g$ and $n_R$.

It is interesting to note that, as long as the $K$-matrix is fixed to
be the two-dimensional matrix $\sigma^x$, a helical edge mode of
the $[0,1]$ phase cannot be gapped.
The addition of a helical edge mode of the trivial $[1,1]^{-1}$ phase
leads to the gapping of the edge of the composite phase
$[0,1]\oplus[1,1]^{-1}$ with $K=\sigma^x\oplus(-\sigma^x)$.
Since we define SPT phases as topologically stable phases against
addition of an arbitrary number of trivial phases, we concluded that
the $[0,1]$ phase is trivial.
This situation is somewhat reminiscent of the classification of
three-dimensional free-fermion insulators of class A, in which
there are topologically stable two-band insulators (with a Hopf invariant)
that are reduced to trivial insulators upon addition of extra
trivial band insulators.\cite{Schnyder_classification_free_2008,Moore}

\subsection{The $S=2$ AKLT state as an SPT phase with $Z_{2}\rtimes R$ symmetry}
\label{subsec: AKLT}
\begin{figure}
\begin{center}
\includegraphics[width=8cm,clip]{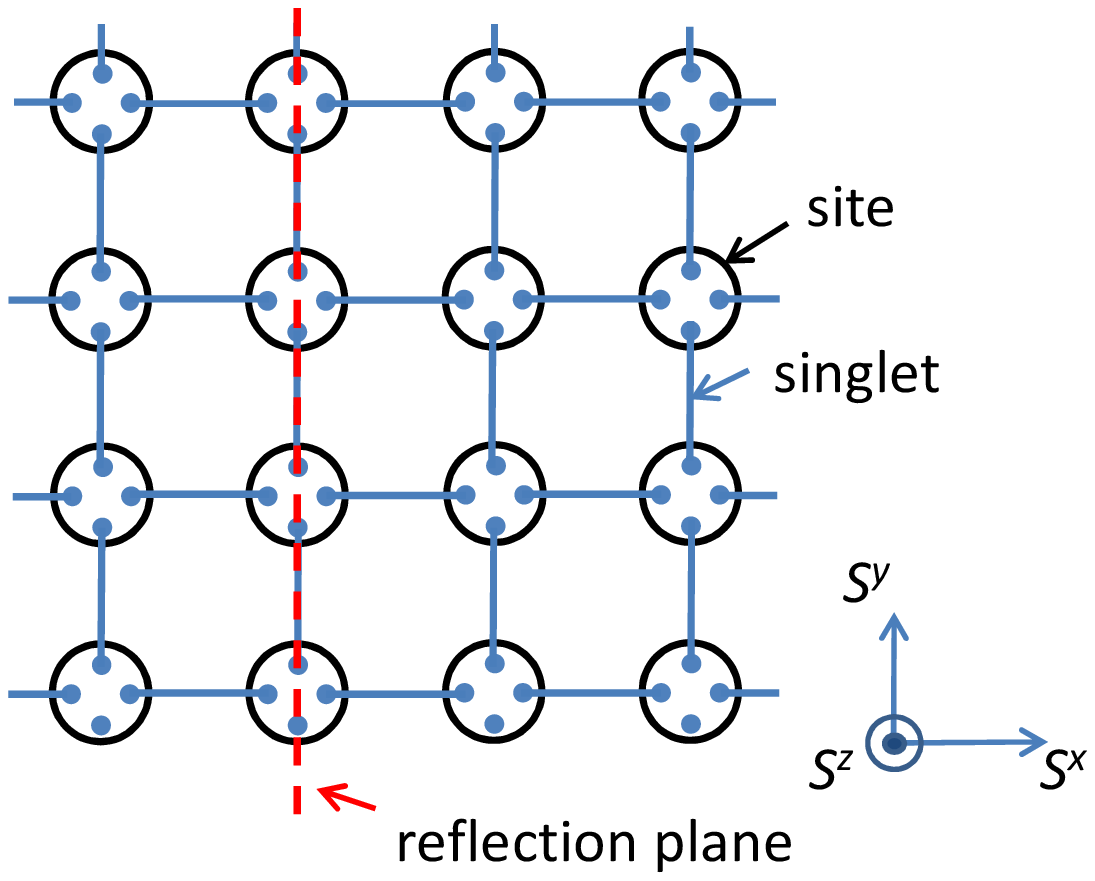}
\end{center}
\caption{
(Color online) Schematic picture of the ground state of
the $S=2$ AKLT model on the square lattice.
Four blue circles on each site represent four $S=1/2$ spins
which form an $S=2$ spin by symmetrization.
Solid blue lines denote spin singlets formed by $S=1/2$ spins
from neighboring sites.
The dashed red line denotes a reflection plane which is perpendicular
to the $x$ direction.
The unpaired $S=1/2$ spins on the edge form
a helical gapless edge mode protected by the symmetry $Z_N \rtimes R$.
}
\label{fig: 2D_AKLT}
\end{figure}

We show that the $S=2$ AKLT model on the 2D square lattice 
realizes a nontrivial SPT phase protected by reflection symmetry $R$
and spin $\pi$-rotation symmetry, $Z_{2}\rtimes R$.
In the $S=2$ AKLT model, $S=2$ spins are placed
on the square lattice and interact with each other
according to the Hamiltonian%
\cite{Affleck_AKLT_honycomb_model1988,Kennedy_2DAKLT_1988}
\begin{eqnarray}
H&=&\sum_{\langle i,j \rangle } P_4(i,j).
\end{eqnarray}
Here,
$P_J(i,j)$ is a projection operator acting on the Hilbert space spanned by
the two spins $\bm{S}_i$ and $\bm{S}_j$ at sites $i$ and $j$
onto the subspace of total spin $J$. 
The summation is taken over all pairs of neighboring sites
$\langle i,j \rangle$.
The AKLT Hamiltonian is written in terms of spin operators as
\begin{eqnarray}
H&=&\sum_{\langle i,j \rangle } \!\!\frac{1}{40320}\prod_{J=0,1,2,3}
 \left[ (\bm{S}_i+\bm{S}_j)^2 - J(J+1)\right] \nonumber \\
 &=&\sum_{\langle i,j \rangle } \left[
\frac{1}{2520}(\bm{S}_i\cdot\bm{S}_j)^4
 +\frac{1}{180}(\bm{S}_i\cdot\bm{S}_j)^3  \right.  \nonumber \\
&&\qquad \left.+\frac{1}{40}(\bm{S}_i\cdot\bm{S}_j)^2
 +\frac{1}{28}(\bm{S}_i\cdot\bm{S}_j) \right] .  
\label{eq: model_spin}
\end{eqnarray}
The ground state of this model is a valence bond solid 
or the AKLT state.
In this state each $S=2$ spin is decomposed into four $S=1/2$ spins, and
at every link a spin singlet is formed by a pair of $S=1/2$ spins from
the sites connected by the link,
as schematically shown in Fig.\ \ref{fig: 2D_AKLT}.
The ground state is obtained by projecting the wave function
built out of $S=1/2$ spins
onto the original Hilbert space spanned by $S=2$ spins.
The bulk excitations are gapped due to the singlet formation.
Along the edge of the square lattice, however,
an unpaired $S=1/2$ spin appears at every site as shown
in Fig.\ \ref{fig: 2D_AKLT}.
These free effective $S=1/2$ spins at the boundary
form dispersionless zero-energy edge states.

Now we are going to show that the $S=2$ AKLT state is in the
SPT phase $[k_g,n_g,n_R]=[1,0,1]$
defined in Eq.~(\ref{eq: g_rt_R trans_law Ug=1}) with $N=2$.
The AKLT model respects the symmetry group $Z_2\rtimes R$. 
The Hamiltonian is invariant under the reflection
with respect to the $x$ direction (which is parallel
to the edge) and under the $\pi$-rotation around
the $S^z$ axis as shown in Fig.\ \ref{fig: 2D_AKLT}. 
Under these transformations the effective $S=1/2$ spins at the boundary
are transformed as
\begin{subequations}
\label{transformations of spins}
\begin{eqnarray}
g 
\begin{pmatrix}
S^x(x) \\ S^y(x) \\ S^z(x)
\end{pmatrix}
g^{-1} 
&=&
\begin{pmatrix}
-S^x(x) \\ -S^y(x) \\ S^z(x)
\end{pmatrix},
\\
R 
\begin{pmatrix}
S^x(x) \\ S^y(x) \\ S^z(x)
\end{pmatrix}
R^{-1}  
&=&
\begin{pmatrix}
S^x(-x) \\ -S^y(-x) \\ -S^z(-x)
\end{pmatrix},
\end{eqnarray}
\end{subequations}
where $g$ denotes the operator for $\pi$-rotation around the $S^z$ axis.
Actually, the group $Z_2\rtimes R$ is identical to $Z_2\times R$.
Here we opted to write $Z_2\rtimes R$ because we utilize results
from Sec.~\ref{subsection: boson_zn_rt_R} and
because the group structure is $Z_N\rtimes R$ for $N>2$ in the case
when the $Z_N$ rotation axis is parallel to the reflection plane
(Fig.~\ref{fig: rot_axis}).
Incidentally, we present the classification of SPT phases protected
by $Z_N\times R$ symmetry in Appendix.

As we mentioned above, the AKLT state has free $S=1/2$ spins
at the boundary that respects the $Z_2\rtimes R$ symmetry.
When the Hamiltonian is perturbed away from the AKLT point
given by Eq.~{(\ref{eq: model_spin})},
we expect that the free spins should interact with each other
antiferromagnetically.
Indeed, this picture is supported by the recent numerical
study\cite{Cirac_ES_AKLT2011,Lou_ES_AKLT2011} 
in which
the entanglement spectrum of the AKLT model on the square
lattice is shown to correspond to the low-energy spectrum of
the spin-1/2 antiferromagnetic Heisenberg chain,
i.e., a conformal field theory with central charge $c=1$.
Thus, the low-energy effective theory for the boundary $S=1/2$
spins below the energy scale of the bulk spin gap should be
the conformal field theory that describes the low-energy excitations of
the spin-$\frac12$ antiferromagnetic Heisenberg model.
In this low-energy theory the boundary $S=1/2$ spins can be
bosonized\cite{giamarchi2003quantum}
\begin{subequations}
\label{eq: model_bosonization}
\begin{eqnarray}
S^z(x)&=&-\frac{1}{2\pi}\partial_x \varphi(x) +A(-1)^x \cos \varphi(x),  \\
S^+(x)&=&e^{-i\vartheta(x)} [B (-1)^x +C \cos \varphi(x) ],
\end{eqnarray}
where $A$, $B$, and $C$ are real constants, and
the bosonic fields $\varphi(x)$ and $\vartheta(x)$ obey
the commutation relation
\begin{equation}
[\varphi(x),\vartheta(x')] =-i\pi +i\pi\mathrm{sgn}(x-x').
\end{equation}
\end{subequations}
Here $\mathrm{sgn}(x)$ equals 1, 0, and $-$1 for $x>0$, $x=0$, and $x<0$,
respectively.
Substituting the bosonization formula (\ref{eq: model_bosonization})
into Eq.~(\ref{transformations of spins}),
we find that the bosonic fields defined by
$\bm{\phi}(x)=(\phi_1(x),\phi_2(x))^T=(-\vartheta(x),\varphi(x))^T$
are transformed as 
\begin{subequations}
\label{transformation of phi}
\begin{eqnarray}
g \bm{\phi}(x) g^{-1} &=& \bm{\phi}(x) +\pi \bm{e}_1, \\
R \bm{\phi}(x) R^{-1} &=& -\sigma^z \bm{\phi}(-x) +\pi \bm{e}_2,
\end{eqnarray}
\end{subequations}
and obey the commutation relation
$[\phi_I(x),\partial_{x'}\phi_J(x')]=2\pi i(\sigma^x)_{IJ}\delta(x-x')$
with $I,J=1,2$.
The transformation of bosonic fields in Eq.\ (\ref{transformation of phi})
corresponds to the one for the SPT phase $[1,0,1]$ that we discussed
in Sec.~\ref{subsection: boson_zn_rt_R}.
Therefore, the gapless edge modes are stable against perturbations
as long as the $Z_2\rtimes R$ symmetry is preserved.
The [1,0,1] phase is a $Z_2$ nontrivial SPT phase.
Since $[1,0,1]\oplus[1,0,1]=0$, two coupled copies of the $S=2$ AKLT model
will have a trivial gapped ground state where edge spins are gapped
by couplings between the copies.

The symmetry protection of edge modes can be intuitively understood
as follows.
In principle, the edge modes can be gapped out
by perturbations that cause singlet formation or magnetization.
However, the ground states gapped by such perturbations  break
the symmetry; singlet formation or out-of-plane magnetization
(magnetization along the $S^z$ axis) breaks the reflection symmetry,
and in-plane magnetization (magnetization along the $S^x$ or $S^y$ axis)
breaks the $\pi$-rotation symmetry.

\section{Discussions}
So far we have studied 2D SPT phases protected by
$Z_N\rtimes R$ symmetry.
The $Z_N$ symmetry can be considered as $2\pi/N$ rotation about a fixed
axis in the spin space in integer-spin systems
(note that $g^N=e$ for any $g\in Z_N$ for integer-$S$ spins).
In this case the semi-direct product group structure $Z_N\rtimes R$
is realized when the spin-rotation axis is parallel to the reflection plane.

As we mentioned at the beginning of Sec.~III,
for integer-spin systems we can also consider other combinations of
reflection symmetry and spin-rotation symmetry:
$Z_N\times R$, $\mathrm{U(1)}\rtimes R$, and $\mathrm{U(1)}\times R$.
The symmetry group $Z_N \times R$ is realized in spin systems
with a spin-rotation axis perpendicular to a reflection plane.
The continuous U(1) rotation symmetry can be viewed as $N\to\infty$ limit of
the discrete $Z_N$ symmetry.
The classification of SPT phases for those symmetries is obtained
in a similar manner as in the case of $Z_N\rtimes R$;
the results are summarized in Table \ref{table: Bose_table}
(For details, see Appendix).

Table \ref{table: Bose_table} indicates that the Abelian group structures
of SPT phases are different depending on whether the spin-rotation
symmetry is discrete $Z_N$ or continuous U(1).
For example, the Abelian group of SPT phases for $\mathrm{U(1)}\rtimes R$
is $\mathbb{Z}_2$, while it is $\mathbb{Z}_2\times\mathbb{Z}_2$
for $Z_N \rtimes R$ with even $N$.
Furthermore, the $Z_N \times R$ symmetry supports SPT phases of $\mathbb{Z}_2\times\mathbb{Z}_2$
while the $\mathrm{U(1)}\times R$ symmetry allows only
the trivial gapped phase without gapless edge modes.
These differences in the Abelian group structures of SPT phases
arise from an additional constraint present in the case of U(1) symmetry.
That is, any U(1) transformation is continuously connected to the identity
transformation as the rotation angle goes to zero; this poses
a further constraint on SPT phases realized in the presence of
the U(1) symmetry compared to the discrete symmetries.

We finish this section with a comment on the effect of disorder
on SPT phases with reflection symmetry.
Considering the stability of surface Dirac fermions in weak topological insulators\cite{Mong_disorderWTI_2012,Ringel_disorderWTI_2012,Fu_disorderWTI_2012,Fulga_disorderWTI_2012,Morimoto_disorderWTI_2014,Obuse_2014}
and topological crystalline insulators,\cite{Fulga_disorderWTI_2012,Morimoto_disorderTCI_2015}
we can expect that edge modes remain gapless and metallic in the presence
of disorder that keeps the reflection symmetry on average.
The disorder effect is experimentally relevant and an important issue.
Detailed analysis is left for a future work.

\section{ACKNOWLEDGMENTS}
This work was partly supported by JSPS KAKENHI Grant Number 15K05141 and
by the RIKEN iTHES Project.

\appendix*
\section{Other bosonic SPT phases}\label{app: bosonic_app}

   \subsection{$\mathrm{U(1)}\rtimes R$ symmetry} 
In this appendix, we obtain the classification of SPT phases
protected by $\mathrm{U(1)}\rtimes R$ symmetry
by making use of the similar analysis for the $Z_N\rtimes R$
case presented in Sec.~\ref{subsubsection: boson_zn_rt_R even}.
The obtained result agrees with the one reported in
Ref.~\onlinecite{Hsieh_CS_CPT_2014}.

U(1) rotations are generated from an infinitesimal rotation,
which can be thought of as the $N\to\infty$ limit of
the $Z_N$ transformation 
with $n_g=0$ in Eq.\ (\ref{g phi g^-1 even N}).
Thus, the transformation law of the bosonic fields
$\bm{\phi}:=(\phi_1,\phi_2)^T$ under U(1) rotation by a finite angle $\theta$
is given by
\begin{subequations}
\begin{equation}
u_\theta \bm{\phi} u^{-1}_\theta =
\bm{\phi}
+\theta
\left(
\begin{array}{c}
k \\
0
\end{array}
\right).
\end{equation}
Here $u_\theta$ is an element of the U(1) group whose rotation angle is
$\theta$, and $k$ is an integer.
As in the $Z_N\rtimes R$ case, the transformation under reflection $R$ is
\begin{equation}
R \bm{\phi} R^{-1} = -\sigma^z\bm{\phi} +\pi
\left(
\begin{array}{c}
0  \\
n_R
\end{array}
\right)
\label{eq: U(1)_rt_R}
\end{equation}
\end{subequations}%
with $n_R=0,1$.
As in Sec.~\ref{subsubsection: boson_zn_rt_R even},
topological phases are labeled by the integer indices $[k,n_R]$.
Now we can follow the argument employed for proving the properties (a) and (b)
in Sec.~\ref{subsubsection: boson_zn_rt_R even}
with setting $n_g=0$ in the relevant equations.
In this way we find
\begin{eqnarray}
[k,n_R]&=&
\left\{
\begin{array}{ll}
0, & kn_R=\mathrm{even},\\
\mbox{[1,1]}, & kn_R=\mathrm{odd}.
\end{array}
\right. 
\end{eqnarray}

Next, we prove that the phase $[1,1]$ is not connected to the trivial phase
by showing that the edge modes of $[\bar{k},n_R]\oplus[\bar{k}',n'_R]^{-1}$,
where $\bar{k},\bar{k}',n_R,n_R'=0$ or $1$,
can be gapped out without symmetry breaking only when
$(\bar{k},n_R)=(\bar{k}',n'_R)$ or $\bar{k}=\bar{k}'=n_R=0$.
For this purpose, we need to see how the elementary bosonic fields
$\bm{v}_1\cdot\bm{\phi}=p\phi_1+q\phi_3$ and
$\bm{v}_2\cdot\bm{\phi}=q\phi_2+p\phi_4$ are transformed,
where $p$ and $q$ are coprime integers, $\mathrm{gcd}(p,q)=1$.
In the phase $[\bar{k},n_R]\oplus[\bar{k}',n'_R]^{-1}$,
the fields are transformed as
\begin{subequations}
\begin{eqnarray}
u_\theta (\bm{v}_1\cdot \bm{\phi}) u^{-1}_\theta&=&
\bm{v}_1\cdot \bm{\phi} +\theta (p\bar{k}+q\bar{k}'),  \\
R (\bm{v}_1\cdot \bm{\phi}) R^{-1}&=& -\bm{v}_1\cdot \bm{\phi},  \\
u_\theta (\bm{v}_2\cdot \bm{\phi}) u^{-1}_\theta&=&
\bm{v}_2\cdot \bm{\phi},  \\
R (\bm{v}_2\cdot \bm{\phi}) R^{-1}&=&
-\bm{v}_2\cdot \bm{\phi} +\theta (qn_R+pn'_R).
\end{eqnarray}
\end{subequations}
If $(p,q)=(\mathrm{odd},\mathrm{odd})$, then the edge modes can be gapped out
without symmetry breaking only when $(\bar{k},n_R)=(\bar{k}',n'_R)$.
If $(p,q)=(\mathrm{even},\mathrm{odd})$, the edge modes can be gapped out
without symmetry breaking only when $\bar{k}=\bar{k}'=n_R=0$.

From the above arguments, we conclude that the Abelian group of SPT phases
protected by U(1)$\rtimes R$ is $\mathbb{Z}_2$ generated by $[1,1]$.

         \subsection{$Z_N \times R$} \label{subsection: boson_zn_t_R} 
In this section
we classify 2D bosonic SPT phases protected by the symmetry group
$G=Z_N \times R$, using the same approach as in
Sec.\ \ref{subsection: boson_zn_rt_R}.

The invariance of the Lagrangian (\ref{S^int_edge}) of edge modes
under symmetry transformation imposes
the transformation rules of $\bm{\phi}$
in Eqs.~(\ref{R transformation_1}) and
(\ref{g transformation_1}).
As in Sec.~\ref{subsection: boson_zn_rt_R},
we take the representation
\begin{equation}
U_R=-\sigma^z,\qquad U_g=\1
\end{equation}
for the $K$-matrix $K=\sigma^x$.
The generators $R$ and $g$ of the symmetry group $G=Z_N\times R$ obey
the relation $R^2=e=g^N$, where $e$ denotes the identity element of $G$.
Accordingly, the representation
$\{U_\mathcal{G},\delta \bm{\phi}_\mathcal{G}\}$ must satisfy
the conditions (\ref{eq: R2=e}) and (\ref{eq: gN=e}).
Furthermore, the algebraic relation $RgRg^{-1}=e$ for the generators
leads to the additional condition
\begin{eqnarray}
&& U^{-1}_gU_RU_gU_R \bm{\phi} + U^{-1}_g (\1+U_RU_g) \delta \bm{\phi}_R
 +U^{-1}_g(U_R-\1) \delta \bm{\phi}_g \nonumber \\
&& =\bm{\phi}, 
\label{eq: ginvRgR=e}
\end{eqnarray}
where we have used the relation
$g^{-1}\bm{\phi}g=U_g^{-1}(\bm{\phi}-\delta\bm{\phi}_g)$.
We discuss cases where $N$ is even and odd separately.

\subsubsection{Even $N$}\label{subsubsection: boson_zn_t_R even} 
It follows from Eqs.~(\ref{eq: R2=e}),\ (\ref{eq: gN=e}), and
(\ref{eq: ginvRgR=e}) that the bosonic fields $\bm{\phi}=(\phi_1,\phi_2)^T$
are transformed as
\begin{subequations}
\label{eq: g_t_R trans_law Ug=1}
\begin{eqnarray}
g \bm{\phi} g^{-1}&=& \bm{\phi} + \pi
\begin{pmatrix}
n_g  \\ 0
\end{pmatrix}
+ \frac{2\pi}{N}
\begin{pmatrix}
0 \\ k_g
\end{pmatrix},  \\
R \bm{\phi} R^{-1}&=& -\sigma^z \bm{\phi} + \pi
\begin{pmatrix}
0 \\ n_R 
\end{pmatrix}
\end{eqnarray}
with
\begin{equation}
n_g,n_R=0,1,
\qquad
k_g=0,\ldots,N-1.
\end{equation}
\end{subequations}
In Eq.~(\ref{eq: g_t_R trans_law Ug=1}b), the phase shift $\delta\phi_1$
under the reflection $R$ is set equal to zero by the basis transformation
in Eq.~(\ref{eq: scheme_trans_gauge}) with $X=\1$ and $\delta\phi_1$
chosen appropriately.

As in Sec.~\ref{subsubsection: boson_zn_rt_R even},
we label SPT phases by the set of integers $[n_g,k_g,n_R]$
that appear in the transformation (\ref{eq: g_t_R trans_law Ug=1}).
We show below that the SPT phases form an Abelian group $\mathbb{Z}_2\times\mathbb{Z}_2$
by proving the following three properties:
\begin{itemize}
 \item[(a)] Phases $[0,k_g,n_R]$ and $[n_g,0,0]$ are trivial
 ($[0,k_g,n_R]=[n_g,0,0]=0$).
 \item[(b)] Any phase is generated from $[1,0,1]$ and $[1,1,0]$,
 which satisfy $[1,0,1]\oplus[1,0,1]=[1,1,0]\oplus[1,1,0]=0$.
 \item[(c)] Two phases $[1,0,1]$ and $[1,1,0]$ are independent generators
 of SPT phases.
\end{itemize}

\textit{Proof of (a):} The null vector condition
[Eq.~(\ref{eq: nullvec_cond_orignal})] with $K=\sigma^x$
allows only pinning potentials of the form,
$\cos(l \phi_1+\alpha_l)$ or $\cos(l \phi_2+\alpha_l)$,
with $l\in\mathbb{Z}$.
When $n_g=0$,
the pinning potential
\begin{equation}
H_{\mathrm{int}}= C\int\! dx_1 \cos(\phi_1)
\end{equation}
is invariant under the transformations and can pin the field $\phi_1$
at $\langle \phi_1\rangle=0$
or $\pi$ depending on the sign of $C$.
No symmetry is broken by pinning.
Thus, $[0,k_g,n_R]$ is connected to a trivial insulator.
When $k_g=n_R=0$, the pinning potential
\begin{eqnarray}
H_{\mathrm{int}}&=& C\int\! dx_1 \cos(\phi_2+\alpha)
\end{eqnarray}
is invariant under the transformation in Eqs.~(\ref{eq: g_t_R trans_law Ug=1})
and can pin the field $\phi_2$ at $\langle \phi_2+\alpha \rangle=0$
or $\pi$ without symmetry breaking.
Thus, $[n_g,0,0]$ is a trivial insulator.

\textit{Proof of (b):}
We first show the following addition relation of SPT phases:
\begin{subequations}
\label{eq: g_t_R Ug=1 summ}
\begin{eqnarray}
{}[n_g,k_g,n_R]\oplus[n_g,k'_g,n'_R] &=& [n_g,k_g+k'_g,n_R+n'_R], \nonumber \\
 && \label{eq: boson_zn_t_R_3}\\
{}[n_g,k_g,n_R]\oplus[n'_g,k_g,n_R] &=& [n_g+n'_g,k_g,n_R].
\label{eq: boson_zn_t_R_4}
\end{eqnarray}
\end{subequations}
The composition of two phases $[n_g,k_g,n_R]$ and $[n_g,k'_g,n'_R]$ has
bosonic fields $\bm{\phi}=(\phi_1,\phi_2,\phi_3,\phi_4)^T$ and
a $K$-matrix $K=\sigma^x\oplus\sigma^x$.
These fields obey the commutation relations
\begin{equation}
[\phi_I(x_1),\partial_{x'_1}\phi_J(x'_1)] =
2\pi i (\sigma^x\oplus\sigma^x)_{I,J}\delta(x_1-x'_1),
\end{equation}
and are transformed as
\begin{subequations}
\begin{eqnarray}
g \bm{\phi} g^{-1}&=&
\bm{\phi} + \pi n_g (\bm{e}_1+\bm{e}_3)
 +\frac{2\pi}{N}(k_g \bm{e}_2 +k'_g \bm{e}_4), \nonumber \\
&& \\
R \bm{\phi} R^{-1}&=&
-(\sigma^z\oplus\sigma^z) \bm{\phi} + \pi(n_R \bm{e}_2 +n'_R \bm{e}_4),
\end{eqnarray}
where
\begin{equation}
n_g,n_R,n'_R=0,1, \qquad k_g,k'_g=0,\ldots,N-1.
\end{equation}
\end{subequations}
Here, $\bm{e}_j$ ($j=1,\ldots, 4$) denotes the $j$th unit vector
in which $(e_{j})_I=\delta_{j,I}$.
We now make basis transformation and define a new set of bosonic fields
\begin{equation}
\bm{\psi}=(\psi_1,\psi_2,\psi_3,\psi_4)^T
=(\phi_1-\phi_3,\phi_2,\phi_3,\phi_2+\phi_4)^T,
\end{equation}
which have the same $K$-matrix and commutators
\begin{equation}
[\psi_I(x),\partial_{x'} \psi_J(x')]
=2\pi i (\sigma^x \oplus \sigma^x)_{I,J} \delta (x-x').
\end{equation}
In the absence of pinning potentials there are two pairs of gapless
helical edge modes: $(\psi_1,\psi_2)$ and $(\psi_3,\psi_4)$.
A potential of the form
\begin{equation}
H_{\mathrm{int}}= C \int\! dx_1 \cos(\psi_1)
\end{equation}
can pin the field and gap out the edge modes in the $(\psi_1,\psi_2)$
sector without symmetry breaking. 
The helical edge states in the $(\psi_3,\psi_4)$ sector remain gapless
and correspond to the phase $[n_g,k_g+k'_g,n_R+n'_R]$. 
Equation (\ref{eq: g_t_R Ug=1 summ}a) follows.
In a similar way, we obtain Eq.~(\ref{eq: g_t_R Ug=1 summ}b) by 
making basis transformation
\begin{equation}
\bm{\psi}'=(\psi'_1,\psi'_2,\psi'_3,\psi'_4)^T
=(\phi_1+\phi_3,\phi_2,\phi_3,\phi_2-\phi_4)^T
\end{equation}
and adding a potential of the form
\begin{equation}
H_{\mathrm{int}2}= C\int\! dx_1 \cos(\psi'_4+\alpha).
\end{equation}
In this case the $(\psi'_3,\psi'_4)$ sector is gapped and can be discarded. 
The edge states in the remaining $(\psi'_1,\psi'_2)$ sector correspond
to the phase $[n_g+n'_g,k_g,n_R]$,
and thus we obtain Eq.~(\ref{eq: g_t_R Ug=1 summ}b).
We find from Eqs.~(\ref{eq: g_t_R Ug=1 summ}) that 
\begin{subequations}
\begin{eqnarray}
{}[1,0,1]\oplus[1,0,1]&=&[0,0,1]=0, \\
{}[0,1,1]\oplus[0,1,1]&=&[0,1,1]=0,
\end{eqnarray}
\end{subequations}
since phase shifts are defined modulo $2\pi$.
Furthermore, using Eqs.~(\ref{eq: g_t_R Ug=1 summ}) successively,
we can reduce any phase $[n_g,k_g,n_R]$ to four phases:
\begin{eqnarray}
{}[n_g,k_g,n_R]&=&
\left\{
\begin{array}{ll}
0, & (k_gn_g,n_gn_R)=(\mbox{e},\mbox{e}),\\
\mbox{[1,1,0]}, & (k_gn_g,n_gn_R)=(\mbox{o},\mbox{e}),\\
\mbox{[1,0,1]}, & (k_gn_g,n_gn_R)=(\mbox{e},\mbox{o}),\\
\mbox{[1,1,0]}\oplus\mbox{[1,0,1]}, & (k_gn_g,n_gn_R)=(\mbox{o},\mbox{o}),
\end{array}
\right. \nonumber 
\end{eqnarray}
where ``e" and ``o" stand for ``even" and ``odd", respectively.

\textit{Proof of (c):}
We show that two phases $[1,0,1]$ and $[1,1,0]$ are neither equivalent
to each other nor connected to the trivial phase 0.
To this end, we first show that the edge modes of the phase
$[n_g,\bar{k}_g,n_R]\oplus[n'_g,\bar{k}'_g,n'_R]^{-1}$
with $\bar{k}_g,\bar{k}'_g=0,1$ cannot be gapped out completely
unless $(n_g,\bar{k}_g,n_R)=(n'_g,\bar{k}'_g,n'_R)$ or
$(n_g,0,0)=(0,\bar{k}'_g,n'_R)$.
It follows from Eq.~(\ref{eq: inverse element}) that
the phase $[n_g,\bar{k}_g,n_R]\oplus[n'_g,\bar{k}'_g,n'_R]^{-1}$
has edge modes described by the bosonic fields
$\bm{\phi}=(\phi_1,\phi_2,\phi_3,\phi_4)^T$
with a $K$-matrix $K=\sigma^x\oplus(-\sigma^x)$. 
The bosonic fields $(\phi_1,\phi_2)$ and $(\phi_3,\phi_4)$
obey the transformation laws of $[n_g,\bar{k}_g,n_R]$ and
$[n'_g,\bar{k}'_g,n'_R]$, respectively.

Gapping out the edge modes $\bm{\phi}=(\phi_1,\phi_2,\phi_3,\phi_4)^T$
requires two pinning potentials 
$\cos(\bm{l}_1\cdot\bm{\phi}+\alpha_1)$ and
$\cos(\bm{l}_2\cdot\bm{\phi}+\alpha_2)$,
whose integer vectors $\bm{l}_1$ and $\bm{l}_2$ must satisfy
Eqs.~(\ref{eq: nullvec_cond_orignal}) and (\ref{eq: nullvec_cond_symm}),
or equivalently, Eqs.~(\ref{eq: R-inv_null_vec_cond}).
Solutions to these equations are given by
\begin{eqnarray}
\bm{l}_1&=& (\alpha p, \beta q,\alpha q, \beta p)^T, \\
\bm{l}_2&=& (\alpha' p, \beta' q,\alpha' q, \beta' p)^T,
\end{eqnarray}
with $\alpha$, $\beta$, $\alpha'$, $\beta'$, $p$, $q\in \mathbb{Z}$.
If $pq=0$ and $p\neq q$, then the elementary bosonic fields defined
in Eq.~(\ref{eq: v alpha}) are given by
$(\bm{v}_1\cdot \bm{\phi},\bm{v}_2\cdot \bm{\phi})=(\phi_3,\phi_2)$
or $(\phi_1,\phi_4)$.
If $pq\neq 0$, we can assume $\mathrm{gcd}(p,q)=1$ and obtain
the elementary bosonic fields 
$\bm{v}_1\cdot\bm{\phi}=p\phi_1+q\phi_3$ and
$\bm{v}_2\cdot\bm{\phi}=q\phi_2+p\phi_4$.
In either case, the fields are transformed as
\begin{subequations}
\label{eq: g_t_R va vb}
\begin{eqnarray}
g (\bm{v}_1 \cdot\bm{\phi}) g^{-1} &=&
 \bm{v}_1 \cdot\bm{\phi} + \pi(pn_g+qn'_g), \\
R (\bm{v}_1 \cdot\bm{\phi}) R^{-1} &=& -\bm{v}_1 \cdot\bm{\phi}, \\
g (\bm{v}_2 \cdot\bm{\phi}) g^{-1} &=&
 \bm{v}_2 \cdot\bm{\phi} + \frac{2\pi}{N}(q\bar{k}_g+p\bar{k}'_g) , \\
R (\bm{v}_2 \cdot\bm{\phi}) R^{-1} &=&
 \bm{v}_2 \cdot\bm{\phi} + \pi(qn_R^{}+pn'_R),
\end{eqnarray}
\end{subequations} 
where we assume $(p,q)=(1,0)$ or $(0,1)$ if $pq=0$.
When $(p,q)=(\mathrm{odd}, \mathrm{odd})$, the phase shifts
in Eqs.~(\ref{eq: g_t_R va vb}a), (\ref{eq: g_t_R va vb}c),
and (\ref{eq: g_t_R va vb}d) can be equal to zero (mod $2\pi$)
only when $(n_g,\bar{k}_g,n_R)=(n'_g,\bar{k}'_g,n'_R)$.
This means that the edge modes cannot be gapped out
without symmetry breaking unless $(n_g,\bar{k}_g,n_R)=(n'_g,\bar{k}'_g,n'_R)$.
Similarly, when $(p,q)=(\mathrm{even}, \mathrm{odd})$,
the edge modes can be gapped out without symmetry breaking
only if $\bar{k}_g=n_R=n'_g=0$,
i.e., $[n_g,\bar{k}_g,n_R]=[n'_g,\bar{k}'_g,n'_R]=0$.
Thus, the two phases $[1,0,1]$ and $[1,1,0]$ are inequivalent,
and both of them are distinct from the trivial phase.

From (a), (b), and (c), we conclude that, when $N$ is even,
the Abelian group of the SPT phases
protected by $Z_N\times R$ is $\mathbb{Z}_2\times\mathbb{Z}_2$
generated by $[1,0,1]$ and $[1,1,0]$.

\subsubsection{Odd $N$}\label{subsubsection: boson_zn_t_R odd}
When $N$ is odd, the transformation law of the bosonic fields
are determined from Eqs.~(\ref{eq: R2=e}),\ (\ref{eq: gN=e}),
and (\ref{eq: ginvRgR=e}) to be
\begin{subequations}
\label{eq: g_t_R trans_law Ug=1 odd}
\begin{eqnarray}
g \bm{\phi} g^{-1}&=& \bm{\phi} + \frac{2\pi}{N}
\begin{pmatrix}
0 \\ k_g
\end{pmatrix},  \\
R \bm{\phi} R^{-1}&=& -\sigma^z \bm{\phi} + \pi
\begin{pmatrix}
0 \\ n_R
\end{pmatrix},
\end{eqnarray}
with
\begin{equation}
n_R=0,1,
\qquad
k_g=0,\ldots,N-1.
\end{equation}
\end{subequations}
In Eqs.~(\ref{eq: g_t_R trans_law Ug=1 odd}b),
the phase shift $\delta\phi_1$ caused by the reflection $R$
is set equal to zero by the basis transformation
in Eq.~(\ref{eq: scheme_trans_gauge})
with $X=\1$ and appropriately chosen $\Delta\phi_1$.
The phase shift of $\phi_1$ vanishes
in Eq.\ (\ref{eq: g_t_R trans_law Ug=1 odd}a)
because of the conditions (\ref{eq: gN=e}) and (\ref{eq: ginvRgR=e}),
i.e., $g^N=e$ and $RgRg^{-1}=e$.

No SPT phase is allowed in this case because
$\phi_1$ acquires no phase shift under the symmetry transformations.
The potential
\begin{equation}
H_{\mathrm{int}}= C\int\! dx_1 \cos(\phi_1+\alpha_1)
\label{H_int odd N}
\end{equation}
is invariant under the transformations in
Eqs.~(\ref{eq: g_t_R trans_law Ug=1 odd}) and can pin the field $\phi_1$
at $\langle \phi_1+\alpha \rangle=0$ or $\pi$ depending on the sign of $C$.
No symmetry is broken by the pinning.
Thus, there is only a topologically trivial gapped phase.

         \subsection{$\mathrm{U(1)}\times R$}
The classification of SPT phases under U(1)$\times R$ symmetry
is obtained by taking the limit $N\to \infty$ in the classification
for the $Z_N\times R$ symmetry. 
The result is in agreement with Ref.~\onlinecite{Hsieh_CS_CPT_2014}. 
Under the U(1) rotation by a finite angle $\theta$
the bosonic fields $\bm{\phi}=(\phi_1,\phi_2)^T$ are transformed as
\begin{equation}
u_\theta\bm{\phi}u_\theta^{-1}=
\bm{\phi}
+\theta \begin{pmatrix}0\\ k\end{pmatrix},
\end{equation}
which is obtained by replacing $2\pi/N$ with $\theta$ in
Eq.\ (\ref{eq: g_t_R trans_law Ug=1 odd}a)
[or further setting $n_g=0$ in Eq.\ (\ref{eq: g_t_R trans_law Ug=1}a)].
The transformation under reflection is given by
Eq.\ (\ref{eq: g_t_R trans_law Ug=1 odd}b).
Obviously we can employ the same argument as the one for
the $Z_N\times R$ symmetry with odd $N$.
The potential in Eq.\ (\ref{H_int odd N}) can pin the $\phi_1$ field,
and the resulting gapped state is a trivial state.



%

%
\end{document}